\documentclass[twocolumn,showpacs,preprintnumbers,amsmath,amssymb]{revtex4}
\usepackage{graphicx}
\usepackage{dcolumn}
\usepackage{bm}

\begin{document}
\title{A loophole in Cabello's proof to Bell's theorem without inequalities}
\author{W. LiMing}
\email{wliming@scnu.edu.cn}
\author{Z. L. Tang}
\affiliation{Department of Physics, South China Normal University,
Guangzhou 510631,China.}
\author{C. Liao}
\affiliation{School for Information and Optoelectronic Science and
Engineering, South China Normal University, Guangzhou 510631,
China}
\date{\today}
\begin{abstract}
A loophole in the proofs of Bell's theorem without inequalities is
pointed out. The assumption in these proofs to EPR's physical
reality does not fit EPR's original arguments.
\end{abstract}

\maketitle
The nonlocality of quantum mechanics has been proved
without using inequalities by a few
authors\cite{Cabello,Hardy,GHZ}. However, there exists a common
loophole in these proofs. Take Cabello's proof as an example. In
his proof\cite{Cabello}, Cabello considered two pairs of maximally
entangled particles, (1,2) and (3,4),
\begin{equation}
|\Psi\rangle_{1234}=|\Psi^-\rangle_{12}\otimes|\Psi^-\rangle_{34},
\end{equation}
where
\begin{equation}
|\Psi^-\rangle_{ij} = \frac{1}{\sqrt{2}}
(|0\rangle_i\otimes|1\rangle_j - |1\rangle_i \otimes|0\rangle_j).
\end{equation}
Two observers, Alice and Bob, at a space-like separation have
access to particles (1,3) and particles (2,4), respectively. From
this state one can predict the values of measurement on $\sigma_x$
and $\sigma_z$ of the four particles with the following
probabilities
\begin{eqnarray}
\label{A13}
P(B_2=B_4|A_1A_3=+1)&=&1, \\
\label{a13}
P(b_2=b_4|a_1a_3=+1)&=&1, \\
\label{B24}
P(A_1=a_3|B_2b_4=+1)&=&1, \\
P(a_1=-A_3|b_2B_4=-1)&=&1, \label{b24}\\
P(A_1A_3=+1,a_1a_3=+1&,& \nonumber\\
B_2b_4=+1,b_2B_4=-1)&=&1/8, \label{A1234}
\end{eqnarray}
where $a,A$ are the values measured by Alice on $\sigma_x$ and
$\sigma_z$, respectively, $b,B$ are those by Bob, and subscripts
denote particles. Then Cabello found a contradiction under the
condition that EPR's elements of physical reality\cite{EPR} exist.
Since $\sigma_{z1}\sigma_{z3},
\sigma_{x1}\sigma_{x3},\sigma_{z2}\sigma_{x4},\sigma_{x2}\sigma_{z4}$
are commutative operators, they can be measured simultaneously.
Thus Alice and Bob make joint measurements on these operators.
Cabello found, for a run of measurement which gives
$A_1A_3=+1,a_1a_3=+1, B_2b_4=+1,b_2B_4=-1$, the elements of
physical reality of $\sigma_x$ and $\sigma_z$ of the four
particles contradict each other.

The loophole comes from the fact that (\ref{A13}-\ref{b24}) do not
work together with  (\ref{A1234}). Consider  (\ref{A13}) and
(\ref{A1234}). From (\ref{A13}), Alice predicts $B_2=B_4$ if she
obtained $A_1A_3=1$ by measuring $\sigma_{z1}\sigma_{z3}$ on her
particles (1,3). However, this prediction won't be agreed by Bob,
because Bob had obtained $B_2b_4,b_2B_4$ by measuring
$\sigma_{z2}\sigma_{x4}$ and $\sigma_{x2}\sigma_{z4}$ on his
particles (2,4). Since $[\sigma_{z2},\sigma_{x2}\sigma_{z4}] \ne
0$, $[\sigma_{z4},\sigma_{z2}\sigma_{x4}] \ne 0$, Bob has no idea
for the values of $\sigma_{z2}$ and $\sigma_{z4}$. Especially, it
is impossible to test Alice's prediction after Bob has measured
$\sigma_{z2}\sigma_{x4}$ and $\sigma_{x2}\sigma_{z4}$. Alice's
prediction has been disturbed by Bob's measurement. Therefore,
Alice's prediction is not true for the joint measurement of Alice
and Bob.

According to EPR\cite{EPR}, an element of physical reality exists
only when a quantity can be predicted with certainty and without
disturbing the system. In Alice and Bob's joint measurement they
in fact cannot predict anything consistently, or Alice's
prediction for particles (2,4) has been disturbed by Bob's
measurement, thus elements of physical reality in the sense of EPR
cannot be assumed.  Therefore, Cabello's final contradiction does
not make sense. This is the loophole.

If a contradiction for the elements of physical reality of
$B_2,B_4,b_2,b_4$ from (\ref{A13},\ref{a13}) and
\begin{equation}
P(A_1A_3=+1,a_1a_3=+1)=1/4,
\end{equation}
could be found, that would be a true disaster of EPR's physical
reality.

The same problem exists between (\ref{a13}) and (\ref{A1234}),
(\ref{B24}) and (\ref{A1234}), and (\ref{b24}) and (\ref{A1234}).

This loophole also exists in GHZ nonlocality\cite{GHZ},Hardy's
argument\cite{Hardy}.

This work was supported by the National Fundamental Research
Program under Grant No 2001CD309300.

\end{document}